%% file: paper.tex
\documentclass[english,nofootinbib]{article}
\usepackage[T1]{fontenc}
\usepackage[latin9]{inputenc}
\usepackage{geometry}
\usepackage{color}
\usepackage{mathtools}
\usepackage{amsmath}
\usepackage{amsthm}
\usepackage{amssymb}

\makeatletter

\providecommand{\tabularnewline}{\\}

\numberwithin{equation}{section}
\numberwithin{figure}{section}
\newcommand{\lyxaddress}[1]{
\par {\raggedright #1
\vspace{1.4em}
\noindent\par}
}

\usepackage{tikz}
\usetikzlibrary{decorations.markings,positioning}
\usepackage{url}
\usepackage{hyperref}
\usepackage{mathtools}
\usepackage{cite}

\newcommand{\myraise}[1]{%
    \mathrel{\raisebox{20pt}{$#1$}}%
}%

\def\llltwocell{\supermorphism {\xy@ {start of 2-cell}{\begingroup }\twocell@ }[0,-3]{}}
\def\rrrtwocell{\supermorphism {\xy@ {start of 2-cell}{\begingroup }\twocell@ }[0,+3]{}}

\makeatother

\usepackage{babel}
\begin{document}

\title{2-connections, a lattice point of view}

\author{B. Bouzid\thanks{bouzid.badreddine@gmail.com}\hspace{0.5em}and
M. Tahiri\date{}}
\maketitle

\lyxaddress{\begin{center}
\textit{Laboratoire de Physique Théorique d\textquoteright Oran (LPTO),
Université d\textquoteright Oran, BP 1524 El M\textquoteright Naouer,
31100 Es-Senia, Oran, Algeria}
\par\end{center}}
\begin{abstract}
We show that the transition laws for a 2-connection can be recovered
by discretizing the base 2-space of a 2-bundle into an Euclidean hypercubic
lattice. The aim of this work is to serve as an example of how important
results in higher gauge theory, which have been derived in a continuous
setting, can also be derived in the lattice scheme.
\end{abstract}

\section{Introduction}

In ordinary gauge theory built upon a $G$-principal bundle $E\xrightarrow{p}B$,
a connection $A$ describes parallel transport of point particles
along paths. This connection can be locally seen as a $Lie(G)$-valued
1-form on the base space $B$, hence it associates a group element
$\mathsf{hol}_{A}(\gamma)\in G$ to each path $\gamma$ of the space,
called the holonomy of $A$ along $\gamma$. In this way, group elements
become associated to paths of the space. We call a configuration or
coloring of this space a given choice of such associations. In lattice
gauge theory, the base space is discretized into an Euclidean hypercubic
lattice with lattice spacing $a$, physical laws are recovered by
taking the limit $a\to0$ (see e.g. \cite{makeenko2002methods} and
references therein).

On the other hand, transformations of extended objects like strings
cannot be described using such a connection, since strings move along
surfaces, whereas point particles move along paths, gauge theories
need to be extended to include connections that can describe parallel
transport of both point particles and strings along paths and surfaces
respectively. There will certainly be some interplay between these
two kinds of transformations, and this should be handled by the extended
theory. This theory is called higher gauge theory \cite{baez2011invitation}
and is the extension of gauge theory in the langage of higher category
theory, which is well suited to deal with such problems. Higher gauge
theory is based on generalizations of spaces, groups, bundles and
connections to, respectively, 2-spaces, 2-groups, 2-bundles and 2-connections
using the so-called enrichment and internalization processes.

Our main goal is to give a simple example of application of lattice
technics to higher gauge theory in order to recover the transformation
laws for a 2-connection. Although these laws have already been derived
in a continuous setting \cite{schreiber2011smooth,schreiber2005loop},
formulating higher gauge theory on a lattice has its own benefit:
it may be applicable to computer-based numerical simulations (see
e.g. \cite{lipstein2014lattice} and references therein). Let us recall
that passing to the lattice formalism has proven to be a crucial step
for the computer-based numerical simulations of gauge theories in
the past (see e.g. \cite{kawamoto20004} and references therein).

Higher lattice gauge theory involves associating not only group elements
to links of the lattice, but also to its plaquettes. Then, using relations
inherited from higher category theory, important results that lie
at the heart of higher gauge theory can be recovered. In the present
paper, we use some of the techniques that appeared in \cite{girelli2004higher}, but which have initially been introduced in \cite{grosse2001duals}. However, we perform a slight modification in the aforementioned techniques. First, we note that point-wise transformations
have been represented by the authors as if they were propagating over
the lattice. We remark that this choice leads to a trucated form of
the transformation laws. This is mainly due to the higher order $\varepsilon$
terms that vanish, where $\varepsilon$ is what they called the ``height''
of the point-wise transformation. Instead, we describe transformations
propagating on the space (such as the holonomies) by assigning group
elements to links and plaquettes, wherease point-wise transformations
are described by the assignment of group elements to vertices.

\section{Definitions and notations}

We descretize the trivial smooth base 2-space $B$ into an Euclidean
lattice $B=a\mathbb{Z}\times a\mathbb{Z}\times\dots\times a\mathbb{Z}$.
Let $e_{\mu}$ denotes the unit vector in the $\mu$ direction, the
vector $ae_{\mu}$ will be written $\mu$ for short. A link $\gamma_{x\rightarrow\mu}$
stands for the oriented path between the ordered pair of points $(x,\,x+\mu)$

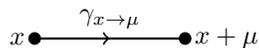
\begin{flushleft}
\input{GammMuNuPath2.tex}
\par\end{flushleft}

whereas a plaquette $\Sigma_{x}^{\mu\rightarrow\nu}$ stands for the
oriented surface with boundary the ordered quadruple $(x,\,x+\mu,\,x+\mu+\nu,\,x+\nu)$

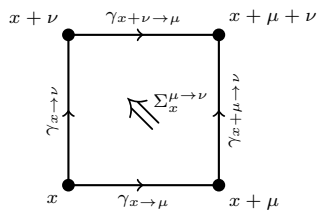
\input{SigmaXYZCell2.tex}

All quantities in the lattice written without argument will be understood
as being evaluated at the origin $x=0$. All groups considered in
this paper are matrix groups.

\begin{flushleft}
The holonomy of the 2-connection for the patch $U_{i}$ along a link
at the origin propagating in the $\mu$ direction $\gamma_{0\rightarrow\mu}$
will be denoted
\[
\mathsf{hol}_{i\mu}=e^{\int_{\gamma_{0\rightarrow\mu}}A_{i}}
\]
\par\end{flushleft}

The holonomy of the 2-connection for the patch $U_{i}$ on a plaquette
at the origin propagating in the $\mu$ and $\nu$ directions $\Sigma_{0}^{\mu\rightarrow\nu}$
will be denoted
\[
\mathsf{hol}_{i\mu\nu}=e^{\int_{\Sigma_{0}^{\mu\rightarrow\nu}}B_{i}}
\]

The 2-group $\mathcal{G}$ will be seen as a 2-category with a single
object denoted $\star$. On each patch $U_{i}$, there is a 2-groupoid
$\mathcal{P}_{2}(U_{i})$ of thin homotopy classes of smooth lazy
paths and surfaces \cite{baez2011invitation}, the holonomy is then
a 2-functor
\[
\mathsf{hol}_{i}:\mathcal{P}_{2}(U_{i})\to\mathcal{G}
\]
that takes each point of $U_{i}$ to the single object $\star$ of
$\mathcal{G}$
\[
\mathsf{hol}_{i}^{(0)}(x)=\star
\]

The 1-morphism map of the holonomy functor 
\[
\mathsf{hol}_{i}^{(1)}:\mathsf{1Mor}(\mathcal{P}_{2}(U_{i}))\to\mathcal{G}^{(1)}
\]
acts on origin-based links of the lattice as

\input{1MorHolMap.tex}

Although the images of all lattice vertices are always $\star$, for
the sake of clarity we will write $\mathsf{hol}_{i}^{(0)}(x)=x$ for
each vertex $x$

\input{1MorHolMapBIS.tex}

But keeping in mind that, if seen as living in $\mathcal{G}$, all
vertices of this diagram are the single object $\star$, while if
it is seen as living in $\mathcal{P}_{2}(U_{i})$, labels on links
are the coloring of the lattice.

In the same spirit, the 2-morphism map of the holonomy functor
\[
\mathsf{hol}_{i}^{(2)}:\mathsf{2Mor}(\mathcal{P}_{2}(U_{i}))\to\mathcal{G}^{(2)}
\]
acts on origin-based plaquettes of the lattice as

\input{2MorHolMap.tex}

\section{2-connections\label{sec:2-connections}}

Let $E\xrightarrow{p}B$ be a $\mathcal{G}$-2-bundle, where $\mathcal{G}$
is some (strict) smooth 2-group corresponding to the Lie crossed module
$(G,H,t,\alpha)$ and $(\mathfrak{g},\mathfrak{h},dt,d\alpha)$ be
its differential crossed module \cite{baez2004hgt,baez2007higher,schreiber2005loop}.
We choose $B$ to be a trivial smooth 2-space equipped with an ordinary
cover $\{U_{i}\}_{i\in I}$ which is hypercubic-wise, i.e., the opens
$U_{i}$ are open hypercubes. This will ensure that no links and no
plaquettes are partially included in some opens $U_{i}$, except perhaps
for their boundaries. But we remark that if an endpoint of a link
does not belong to the patch $U_{i}$ that contains the rest of the
link, the integration is not going to differ much from that using
the whole link. The transition functions on the cover $\{U_{i}\}_{i\in I}$
are $g_{ij}$, $h_{ijk}$ and $k_{i}$. We will also restrict ourselves
to the case $k_{i}=\mathsf{1}$. On each patch $U_{i}$ the local
holonomy 2-functor $\mathsf{hol}_{i}$ is specified by two differential
forms
\begin{eqnarray*}
A_{i} & \in & \Omega^{1}(U_{i},\mathfrak{g})\\
B_{i} & \in & \Omega^{2}(U_{i},\mathfrak{h})
\end{eqnarray*}
such that the fake curvature vanishes 
\[
F_{A_{i}}+dt(B_{i})=0
\]
where $F_{A_{i}}$ is the curvature 2-form of $A_{i}$.

The transition pseudonatural isomorphism $g_{ij}:\mathsf{hol}_{i}\Rightarrow\mathsf{hol}_{j}$
is specified by the transition functions $g_{ij}$ together with differential
forms $a_{ij}\in\Omega^{1}(U_{ij},\mathfrak{h})$, whereas the modification
$h_{ijk}:g_{ij}g_{jk}\Rrightarrow g_{ik}$ is specified by the transition
functions $h_{ijk}$, such that on every double overlap $U_{ij}$
the following transformation laws hold
\begin{eqnarray}
A_{i} & = & g_{ij}A_{j}g_{ij}^{-1}+g_{ij}\mathsf{d}g_{ij}^{-1}-\text{d}t(a_{ij})\label{eq:1}\\
B_{i} & = & \alpha(g_{ij})(B_{j})+\text{d}a_{ij}+a_{ij}\wedge a_{ij}+\text{d}\alpha(A_{i})\wedge a_{ij}\label{eq:2}
\end{eqnarray}
and on every triple overlap $U_{ijk}$ the following transformation
law holds
\begin{equation}
a_{ij}+\alpha(g_{ij})a_{jk}=h_{ijk}^{-1}a_{ik}h_{ijk}+h_{ijk}^{-1}\text{d}h_{ijk}+h_{ijk}^{-1}\text{d}\alpha(A_{i})h_{ijk}\label{eq:3}
\end{equation}

Let us recall that these transformation laws have already been derived
in \cite{schreiber2005loop,schreiber2011smooth} using a continuous
setting. Here we use a different approach which has been inspired
from \cite{girelli2004higher}. There, the authors used lattice calculus
to recover the fake curvature connection (which has also been previously
derived in a continuous setting) as well as other important results.\textcolor{red}{{} }

\section{Transition laws for the 2-connection}

In higher lattice gauge theory, transition functions are represented
not only by 1-morphisms
\[
\forall x\in\mathsf{Ob}(\mathcal{P}_{2}(U_{ij})):g_{ij}(x)\in\mathcal{G}^{(1)}
\]
but also by 2-morphisms
\[
\forall\gamma\in\mathsf{1Mor}(\mathcal{P}_{2}(U_{ij})):g_{ij}(\gamma)\in\mathcal{G}^{(2)}
\]
such that if $\gamma:x\to y$, we have
\[
g_{ij}(\gamma):\mathsf{hol}_{i}(\gamma)\ g_{ij}(y)\Rightarrow g_{ij}(x)\ \mathsf{hol}_{j}(\gamma)
\]
To derive equation (\ref{eq:1}) we take a link $\gamma_{0\to\mu}\in U_{ij}$,
its images in $\mathcal{G}$ via the 2-connections $(A_{i},B_{i})$
and $(A_{j},B_{j})$ are related by the transition functions as follows

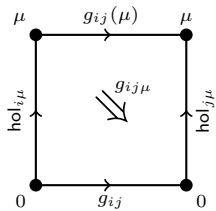
\begin{center}
\input{ElementaySquare2Gauge.tex}
\par\end{center}

The 2-morphism $g_{ij\mu}$ has $\mathsf{hol}_{i\mu}g_{ij}(\mu)$
as source 1-morphism and $g_{ij}\mathsf{hol}_{j\mu}$ as target 1-morphism,
we have then the following relation which stems directly from higher
category theory \cite{baez2011invitation}
\[
g_{ij}\mathsf{hol}_{j\mu}=t(g_{ij\mu})\ \mathsf{hol}_{i\mu}g_{ij}(\mu)
\]
thus $\mathsf{hol}_{i\mu}$ is
\begin{eqnarray*}
\mathsf{hol}_{i\mu} & = & t(g_{ij\mu})^{-1}\ g_{ij}\ \mathsf{hol}_{j\mu}\ g_{ij}(\mu)^{-1}\\
e^{\int_{\gamma_{0\rightarrow\mu}}A_{i}} & = & t(g_{ij\mu}^{-1})\ g_{ij}\ e^{\int_{\gamma_{0\rightarrow\mu}}A_{j}}\ g_{ij}(\mu)^{-1}
\end{eqnarray*}
The differential forms $a_{ij}$ describe the transition pseudonatural
isomorphisms $g_{ij}$ at the plaquette level, thus
\[
g_{ij\mu}=e^{aa_{ij\mu}}
\]
Hereafter, as $a\to0$ the symbol $\approx$ means that we approximate
the equalities by neglecting terms of order higher than the dimension
we are working on.

As $a\to0$ the connection can be considered as constant along each
link, so that we get
\[
e^{\int_{\gamma_{0\rightarrow\mu}}A_{i}}\approx e^{aA_{i\mu}}
\]
On the other hand, using a Taylor expansion in $g_{ij}(\mu)$ and
the derivative of $t$ we finally get
\begin{eqnarray*}
e^{aA_{i\mu}} & \approx & e^{-a\text{d}t(a_{ij\mu})}\ g_{ij}\ e^{aA_{j\mu}}\ (g_{ij}^{-1}+a\partial_{\mu}g_{ij}^{-1})
\end{eqnarray*}
Again, using Taylor expansions of exponentials, we get
\begin{eqnarray*}
1+aA_{i\mu} & \approx & (1-a\text{d}t(a_{ij\mu}))g_{ij}(1+aA_{j\mu})(g_{ij}^{-1}+a\partial_{\mu}g_{ij}^{-1})\\
 & \approx & 1+a\left(g_{ij}\partial_{\mu}g_{ij}^{-1}+g_{ij}A_{j\mu}g_{ij}^{-1}-\text{d}t(a_{ij\mu})\right)
\end{eqnarray*}
 We thus recover (\ref{eq:1}).

Now to derive equation (\ref{eq:2}) we take a plaquette at the origin,
its images in $\mathcal{G}$ via the 2-connections $(A_{i},B_{i})$
and $(A_{j},B_{j})$ are related by the transition functions as follows

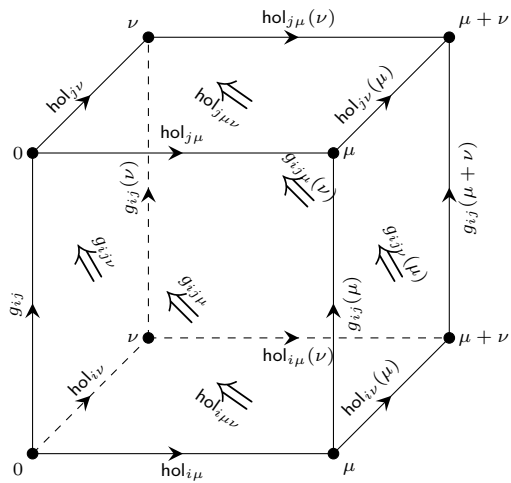
\begin{center}
\input{ColoredCube.tex}
\par\end{center}

The coloring 2-morphism $\mathsf{hol}_{i\mu\nu}$ has source $\mathsf{hol}_{i\mu}\mathsf{hol}_{i\nu}(\mu)$
and target $\mathsf{hol}_{i\nu}\mathsf{hol}_{i\mu}(\nu)$, it sweeps
the bottom side of the cube. Alternatively, it can also be seen as
sweeping the remaining sides of the cube since the diagram commutes
in $\mathcal{G}$.

Let us denote horizontal (vertical) composition of 2-morphisms by
$\circ_{h}$ ($\circ_{v}$).

We note that the horizontal composition of a 2-morphism with a 1-morphism
is a shortcut of the horizontal composition of this 2-morphism with
identity 2-morphism of the 1-morphism, that is, for example for $g\in G$
and $h\in H$ 
\[
h\circ_{h}g\coloneqq h\circ_{h}\mathsf{1}_{g}
\]

We remark first that

\begin{center}
$\myraise{\mathsf{hol}_{i\mu\nu}\circ_{h}g_{ij}(\mu+\nu)=}$\input{cube_proof_9.tex}$\myraise{=\ }$\input{cube_proof_10.tex}$\myraise{=\mathsf{hol}_{i\mu\nu}}$
\par\end{center}

Now to write down the other expression of that 2-morphism (i.e. when
it sweeps the remaining sides), we will need the following pieces
of the cube

\begin{tabular}{r||c||ll||l}
\multicolumn{3}{r}{$\myraise{\mathsf{hol}_{\text{i}\mu}\circ_{h}g_{ij\nu}(\mu)=}$\input{cube_proof_6.tex}$\myraise{=}$\input{cube_proof_7.tex}} & \multicolumn{2}{l}{$\myraise{=\mathsf{hol}_{\text{i}\mu}\rhd g_{ij\nu}(\mu)}$}\tabularnewline
\multicolumn{3}{r}{$\myraise{g_{ij\mu}\circ_{h}\mathsf{hol}_{\text{j}\nu}(\mu)=}$\input{cube_proof_3.tex}$\myraise{=}$\input{cube_proof_4.tex}} & \multicolumn{2}{l}{$\myraise{=g_{ij\mu}}$}\tabularnewline
\multicolumn{3}{r}{$\myraise{g_{ij}\circ_{h}\mathsf{hol}_{j\mu\nu}=}$\input{cube_proof_1.tex}$\myraise{=}$\input{cube_proof_2.tex}} & \multicolumn{2}{l}{$\myraise{=g_{ij}\rhd\mathsf{hol}_{j\mu\nu}}$}\tabularnewline
\end{tabular}

\begin{tabular}{r||c||lll}
\multicolumn{3}{r}{$\myraise{g_{ij\nu}\circ_{h}\mathsf{hol}_{j\mu}(\nu)=}$\input{cube_proof_13.tex}$\myraise{=}$\input{cube_proof_14.tex}} & $\myraise{=g_{ij\nu}}$ & $\myraise{\overset{(\bullet)^{-1}}{\longmapsto}}$\input{cube_proof_15.tex}$\myraise{=g_{ij\nu}^{-1}}$\tabularnewline
\multicolumn{3}{r}{$\myraise{\mathsf{hol}_{i\nu}\circ_{h}g_{ij\mu}(\nu)=}$\input{cube_proof_11.tex}$\myraise{=}$\input{cube_proof_12.tex}} & $\myraise{=\mathsf{hol}_{i\nu}\rhd g_{ij\mu}(\nu)}$ & $\myraise{\overset{(\bullet)^{-1}}{\longmapsto}}$\input{cube_proof_12.5.tex}$\myraise{=\mathsf{hol}_{i\nu}\rhd g_{ij\mu}^{-1}(\nu)}$\tabularnewline
\end{tabular}

Using the following equality,

\begin{center}
\input{cube_proof_10.tex}$\myraise{=\ }$\input{cube_proof_7.tex}$\myraise{\circ_{v}}$\input{cube_proof_4.tex}$\myraise{\circ_{v}}$\input{cube_proof_2.tex}$\myraise{\circ_{v}}$\input{cube_proof_15.tex}$\myraise{\circ_{v}}$\input{cube_proof_12.5.tex}
\par\end{center}

\begin{flushleft}
we get 
\begin{eqnarray*}
\mathsf{hol}_{i\mu\nu} & = & \left[\mathsf{hol}_{i\nu}\rhd g_{ij\mu}^{-1}(\nu)\right]g_{ij\nu}^{-1}\left[g_{ij}\rhd\mathsf{hol}_{j\mu\nu}\right]g_{ij\mu}\left[\mathsf{hol}_{\text{i}\mu}\rhd g_{ij\nu}(\mu)\right]\\
e^{\int_{\Sigma_{0}^{\mu\rightarrow\nu}}B_{i}} & = & \alpha(e^{\int_{\gamma_{0\rightarrow\nu}}A_{i}})(g_{ij\mu}^{-1}(\nu))g_{ij\nu}^{-1}\alpha(g_{ij})(e^{\int_{\Sigma_{0}^{\mu\rightarrow\nu}}B_{j}})g_{ij\mu}\alpha(e^{\int_{\gamma_{0\rightarrow\mu}}A_{i}})(g_{ij\nu}(\mu))
\end{eqnarray*}
\par\end{flushleft}

Again, as $a\to0$ the 2-connection can be considered as constant
on each plaquette, so that
\[
e^{\int_{\Sigma_{0}^{\mu\rightarrow\nu}}B_{i}}\approx e^{a^{2}B_{i\mu\nu}}
\]

\begin{flushleft}
Using the derivative of $\alpha$ and a Taylor expansion on $g_{ij\mu}(\nu)$
we get
\[
e^{a^{2}B_{i\mu\nu}}\approx e^{a\mathsf{d}\alpha(A_{i\nu})}(e^{-aa_{ij\mu}-a^{2}\partial_{\nu}a_{ij\mu}})e^{-aa_{ij\nu}}\alpha(g_{ij})(e^{a^{2}B_{j\mu\nu}})e^{aa_{ij\mu}}e^{a\mathsf{d}\alpha(A_{i\mu})}(e^{aa_{ij\nu}+a^{2}\partial_{\mu}a_{ij\nu}})
\]
Expanding exponentials and after some calculations we get
\begin{eqnarray*}
1+a^{2}B_{i\mu\nu} & \approx & (1+a\mathsf{d}\alpha(A_{i\nu}))(1-aa_{ij\mu}-a^{2}\partial_{\nu}a_{ij\mu})(1-aa_{ij\nu})\\
 &  & \ \ \ \ \alpha(g_{ij})(1+a^{2}B_{j\mu\nu})(1+aa_{ij\mu})(1+a\text{d}\alpha(A_{i\mu}))(1+aa_{ij\nu}+a^{2}\partial_{\mu}a_{ij\nu})\\
 & \approx & 1+a^{2}[\alpha(g_{ij})(B_{j\mu\nu})+\partial_{\mu}a_{ij\nu}-\partial_{\nu}a_{ij\mu}+a_{ij\mu}a_{ij\nu}-a_{ij\nu}a_{ij\mu}\\
 &  & \ \ \ \ +\mathsf{d}\alpha(A_{i\mu})(a_{ij\nu})-\mathsf{d}\alpha(A_{i\nu})(a_{ij\mu})]
\end{eqnarray*}
where we have dropped out the symmetric terms. So we recover (\ref{eq:2}).
\par\end{flushleft}

It is worth noting that in \cite{girelli2004higher} the authors wrote
pointwise gauge transformations as they were propagating in an $\alpha$
direction over a link of length $\varepsilon$. However, it turns
out that with such a choice, transformations like $g_{ij\mu}(\nu)$
would have been
\[
g_{ij\mu}(\nu)\approx e^{\varepsilon aa_{ij\mu}+\varepsilon a^{2}\partial_{\nu}a_{ij\mu}}
\]
and then, the $a_{ij}\wedge a_{ij}$ term would have disappeared from
the transformation laws because of their $a^{2}\varepsilon^{2}$ order.

Finally to derive equation (\ref{eq:3}) we take the triangle that
represents the action of the modification $h_{ijk}$ on transition
functions, its images in $\mathcal{G}$ via the 2-connections $(A_{i},B_{i})$
and $(A_{j},B_{j})$ are related by the transition functions as follows

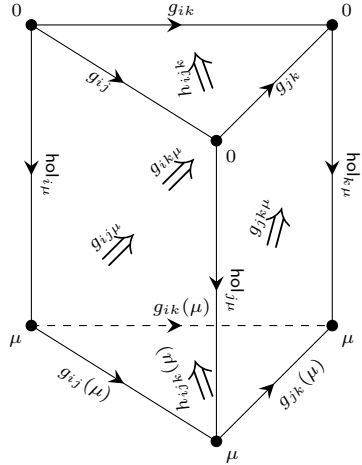
\begin{center}
\input{ColoredPrism.tex}
\par\end{center}

The coloring 2-morphism $g_{ik\mu}$ has source $\mathsf{hol}_{i\mu}\ g_{ik}(\mu)$
and target $g_{ik}\ \mathsf{hol}_{k\mu}$, it sweeps the backside
face of the prism. Alternatively, it can be seen as sweeping the remaining
sides of the prism since the diagram commutes in $\mathcal{G}$. Let
us repeat the previous steps for this diagram.

We remark first that

\begin{center}
$\myraise{g_{ik\mu}=\ }$\input{prism_proof_1.tex}
\par\end{center}

Now we write down the other expression of that 2-morphism

\begin{tabular}{r||c||lll}
\multicolumn{3}{r}{$\myraise{\mathsf{hol}_{i\mu}\circ_{h}h_{ijk}(\mu)=\ }$\input{prism_proof_2.tex}$\myraise{=\ }$\input{prism_proof_3.tex}} & $\myraise{=\mathsf{hol}_{i\mu}\rhd h_{ijk}(\mu)}$ & $\myraise{\overset{(\bullet)^{-1}}{\longmapsto}}$\input{prism_proof_4.tex}$\myraise{=\mathsf{hol}_{i\mu}\rhd h_{ijk}^{-1}(\mu)}$\tabularnewline
\multicolumn{3}{r}{$\myraise{g_{ij\mu}\circ_{h}g_{jk}(\mu)=\ }$\input{prism_proof_9.tex}$\myraise{=\ }$\input{prism_proof_10.tex}} & \multicolumn{2}{l}{$\myraise{=g_{ij\mu}}$}\tabularnewline
\multicolumn{3}{r}{$\myraise{g_{ij}\circ_{h}g_{jk\mu}=\ }$\input{prism_proof_5.tex}$\myraise{=\ }$\input{prism_proof_6.tex}} & \multicolumn{2}{l}{$\myraise{=g_{ij}\rhd g_{jk\mu}}$}\tabularnewline
\end{tabular}

\begin{tabular}{r||c||ll||l}
\multicolumn{3}{r}{$\myraise{h_{ijk}\circ_{h}\mathsf{hol}_{k\mu}=\ }$\input{prism_proof_7.tex}$\myraise{=\ }$\input{prism_proof_8.tex}} & \multicolumn{2}{l}{$\myraise{=h_{ijk}}$}\tabularnewline
\end{tabular}

Using the following equality

\begin{center}
\input{prism_proof_1.tex}$\myraise{=\ }$\input{prism_proof_4.tex}$\myraise{\circ_{v}}$\input{prism_proof_10.tex}$\myraise{\circ_{v}}$\input{prism_proof_6.tex}$\myraise{\circ_{v}}$\input{prism_proof_8.tex}
\par\end{center}

we get
\begin{eqnarray*}
g_{ik\mu} & = & h_{ijk}\left[g_{ij}\rhd g_{jk\mu}\right]g_{ij\mu}\left[\mathsf{hol}_{i\mu}\rhd h_{ijk}^{-1}(\mu)\right]\\
h_{ijk}^{-1}g_{ik\mu}\left[\mathsf{hol}_{i\mu}\rhd h_{ijk}(\mu)\right] & = & \left[g_{ij}\rhd g_{jk\mu}\right]g_{ij\mu}
\end{eqnarray*}
The R.H.S is
\begin{eqnarray*}
\left[g_{ij}\rhd g_{jk\mu}\right]g_{ij\mu} & = & \left(\alpha(g_{ij})g_{jk\mu}\right)g_{ij\mu}\\
 & = & \left(\alpha(g_{ij})e^{aa_{jk\mu}}\right)e^{aa_{ij\mu}}
\end{eqnarray*}
Using Taylor expansions of exponentials we get
\begin{eqnarray*}
\left[g_{ij}\rhd g_{jk\mu}\right]g_{ij\mu} & \approx & \left(\alpha(g_{ij})\left(1+aa_{jk\mu}\right)\right)\left(1+aa_{ij\mu}\right)\\
 & \approx & 1+a\left(a_{ij\mu}+\alpha(g_{ij})a_{jk\mu}\right)
\end{eqnarray*}
The L.H.S is
\[
h_{ijk}^{-1}g_{ik\mu}\left[\mathsf{hol}_{i\mu}\rhd h_{ijk}(\mu)\right]=h_{ijk}^{-1}g_{ik\mu}\left[\alpha(e^{aA_{i\mu}})h_{ijk}(\mu)\right]
\]
Using Taylor expansions of exponentials and of $h_{ijk}(\mu)$, as
well as the derivative of $\alpha$ we get 
\begin{eqnarray}
h_{ijk}^{-1}g_{ik\mu}\left[\mathsf{hol}_{i\mu}\rhd h_{ijk}(\mu)\right] & \approx & h_{ijk}^{-1}\left(1+aa_{ik\mu}\right)\left[\left(1+a\text{d}\alpha(A_{i\mu})\right)\left(h_{ijk}+a\partial_{\mu}h_{ijk}\right)\right]\nonumber \\
 & \approx & 1+a\left(h_{ijk}^{-1}a_{ik\mu}h_{ijk}+h_{ijk}^{-1}\partial_{\mu}h_{ijk}+h_{ijk}^{-1}\text{d}\alpha(A_{i\mu})h_{ijk}\right)\label{lhs}
\end{eqnarray}
This leads us to obtain (\ref{eq:3}).

\section{Conclusion and outlook}

We have shown that calculus on the lattice can systematically be used
to derive important results of higher gauge theory. This is achieved
by coloring plaquettes in addition to links without any other assumption.
The physical laws can be recovered from the $a\to0$ limit. However,
some coherence relations between group elements coloring links and
group elements coloring plaquettes are pivotal for this construction.
These relations prove to be at the heart of the higher category theory:
the first group elements are morphisms (or 1-morphisms) of some 2-group,
while the latter are 2-morphisms between these morphisms. This is
reminiscent of what happens in some gauge theories where symmetries
between symmetries appear due to the presence of second class constraints.
It is worth noting that in $BF$ theory, which is known to have this
kind of metasymmetries, and which has close relations to quantum gravity
\cite{bouzid2012bf}, the gauge fields defining the theory form a
2-connection \cite{cianfrani2011kinematical}.

All this tends to prove that higher category theory is a fertile ground
where theories can be enriched, by systematically extending basic
structures underlying them using the two main tools of higher category
theory: internalization and enrichment \cite{baez2007higher}.

Finally, let us point out that what we have called 2-bundles is also
known under the name of\emph{ gerbes}, and that a whole theory of
differential gerbes already exists \cite{breen2005differential}.
It can be interesting to approach some aspects of this theory using
calculus on lattice defined in the present article, for example, to
found the more general transformation laws when no $k_{i}=1$ restriction
is made.

\bibliographystyle{plain}
\bibliography{ref}

\end{document}

%% file: GammMuNuPath2.tex
\begin{center}
\begin{tikzpicture}[thick, decoration = {
			markings,
			mark=at position 0.5 with {\arrow{>}}
		}]

	\coordinate (0) at (0,0);
	\path[draw, fill=black] (0,0) circle[radius=2pt] node {};

	\coordinate (2) at (2,0);
	\path[draw, fill=black] (2,0) circle[radius=2pt] node {};

	\node at (0) [left] {$x$};
	\node at (2) [right] {$x+\mu$};
	\draw (0) edge[postaction={decorate}] node[above] {$\gamma_{x\rightarrow \mu}$} (2);

\end{tikzpicture}
\end{center}

%% file: SigmaXYZCell2.tex
\begin{center}
\begin{tikzpicture}[thick,decoration={
			markings,
			mark=at position 0.5 with {\arrow{>}}
		}]

	\foreach \x in {0,2} {
		\foreach \y in {0,2} {
			\coordinate (\x\y) at (\x,\y);
			\path[draw, fill=black] (\x,\y) circle[radius=2pt] node {};
		}
	}

	\node at (00) [below left] {\scriptsize $x$};
	\node at (20) [below right] {\scriptsize $x+\mu$};
	\node at (02) [above left] {\scriptsize $x+\nu$};
	\node at (22) [above right] {\scriptsize $x+\mu+\nu$};

	\draw[postaction={decorate}] (00) -- node[below] {\scriptsize $\gamma_{x\rightarrow \mu}$} (20);
	\draw[postaction={decorate}] (02) -- node[auto] {\scriptsize $\gamma_{x+\nu\rightarrow \mu}$} (22);
	\draw[postaction={decorate}] (00) -- node[auto] {\rotatebox{90} {\scriptsize $\gamma_{x\rightarrow \nu}$}} (02);
	\draw[postaction={decorate}] (20) -- node[right] {\rotatebox{90} {\scriptsize $\gamma_{x+\mu\rightarrow \nu}$}} (22);

	\node at (1,1) {\rotatebox{135}{\LARGE{}$\Rightarrow$}};
	\node at (1.5,1.1) {\scriptsize $\Sigma_{x}^{\mu\rightarrow \nu}$};

\end{tikzpicture}
\end{center}

%% file: 1MorHolMap.tex
\begin{center}
\begin{tikzpicture}[thick, decoration = {
			markings,
			mark=at position 0.5 with {\arrow{>}}
		}]

	\coordinate (hol_i) at (-0.8,0);
	\node at (hol_i) {$\mathsf{hol}_{i}^{(1)}(\ \ $};

	\coordinate (0) at (0,0);
	\path[draw, fill=black] (0,0) circle[radius=2pt] node {};

	\coordinate (2) at (2,0);
	\path[draw, fill=black] (2,0) circle[radius=2pt] node {};

	\node at (0) [left] {$0$};
	\node at (2) [right] {$\mu$};
	\draw (0) edge[postaction={decorate}] node[above] {$\gamma_{0\rightarrow\mu}$} (2);

	\coordinate (3) at (3,0);
	\node at (3) {$)\ =\ \ $};

	\coordinate (4) at (4,0);
	\coordinate (6) at (6,0);

	\node at (4) [left] {$\star$};
	\node at (6) [right] {$\star$};

	\draw (4) edge[->] node[above] {$\mathsf{hol}_{i\mu}$} (6);


\end{tikzpicture}
\end{center}

%% file: 1MorHolMapBIS.tex
\begin{center}
\begin{tikzpicture}[thick, decoration = {
			markings,
			mark=at position 0.5 with {\arrow{>}}
		}]

	\coordinate (hol_i) at (-0.8,0);
	\node at (hol_i) {$\mathsf{hol}_{i}^{(1)}($};

	\coordinate (0) at (0,0);
	\path[draw, fill=black] (0,0) circle[radius=2pt] node {};

	\coordinate (2) at (2,0);
	\path[draw, fill=black] (2,0) circle[radius=2pt] node {};

	\node at (0) [left] {$0$};
	\node at (2) [right] {$\mu$};

	\draw (0) edge[postaction={decorate}] node[above] {$\gamma_{0\rightarrow\mu}$} (2);

	\coordinate (3) at (3,0);
	\node at (3) {$)\ \ \ =$};

	\coordinate (4) at (4,0);
	\path[draw, fill=black] (4,0) circle[radius=2pt] node {};

	\coordinate (6) at (6,0);
	\path[draw, fill=black] (6,0) circle[radius=2pt] node {};

	\draw (4) edge[postaction={decorate}] node[above] {$\mathsf{hol}_{i\mu}$} (6);

	\node at (4) [left] {$0$};
	\node at (6) [right] {$\mu$};


\end{tikzpicture}
\end{center}

%% file: 2MorHolMap.tex
\begin{center}
\begin{tikzpicture}[thick,decoration={
			markings,
			mark=at position 0.5 with {\arrow{>}}
		}]

	\coordinate (hol_i) at (-0.8,1);
	\node at (hol_i) {$\mathsf{hol}_{i}^{(2)}$ {\LARGE{}$(\ \ $}};

	\foreach \x in {0,2,4,6} {
		\foreach \y in {0,2} {
			\coordinate (\x\y) at (\x,\y);
			\path[draw, fill=black] (\x,\y) circle[radius=2pt] node {};
		}
	}

	\node at (00) [below left] {\scriptsize $0$};
	\node at (20) [below right] {\scriptsize $\mu$};
	\node at (02) [above left] {\scriptsize $\nu$};
	\node at (22) [above] {\scriptsize $\mu+\nu$};

	\draw[postaction={decorate}] (00) -- node[below] {\scriptsize $\gamma_{0\rightarrow\mu}$} (20);
	\draw[postaction={decorate}] (02) -- node[auto] {\scriptsize $\gamma_{\nu\rightarrow\mu}$} (22);
	\draw[postaction={decorate}] (00) -- node[auto] {\rotatebox{90} {\scriptsize $\gamma_{0\rightarrow\nu}$}} (02);
	\draw[postaction={decorate}] (20) -- node[right] {\rotatebox{90} {\scriptsize $\gamma_{\mu\rightarrow\nu}$}} (22);

	\node at (1,1) {\rotatebox{135}{\LARGE{}$\Rightarrow$}};
	\node at (1.5,1.1) {\scriptsize $\Sigma_{0}^{\mu\rightarrow\nu}$};

	\coordinate (equal) at (2.8,1);
	\node at (equal) {{\LARGE{}$)$} $\ \ =$};

	\node at (40) [below left] {\scriptsize $0$};
	\node at (60) [below right] {\scriptsize $\mu$};
	\node at (42) [above left] {\scriptsize $\nu$};
	\node at (62) [above] {\scriptsize $\mu+\nu$};

	\draw[postaction={decorate}] (40) -- node[below] {\scriptsize $\mathsf{hol}_{i\mu}$} (60);
	\draw[postaction={decorate}] (42) -- node[auto] {\scriptsize $\mathsf{hol}_{i\mu}(\nu)$} (62);
	\draw[postaction={decorate}] (40) -- node[auto] {\rotatebox{90} {\scriptsize $\mathsf{hol}_{i\nu}$}} (42);
	\draw[postaction={decorate}] (60) -- node[right] {\rotatebox{90} {\scriptsize $\mathsf{hol}_{i\nu}(\mu)$}} (62);

	\node at (5,1) {\rotatebox{135}{\LARGE{}$\Rightarrow$}};
	\node at (5.5,1.1) {\scriptsize $\mathsf{hol}_{i\mu\nu}$};

\end{tikzpicture}
\end{center}

%% file: ElementaySquare2Gauge.tex
\begin{center}
\begin{tikzpicture}[thick,decoration={
			markings,
			mark=at position 0.5 with {\arrow{>}}
		}]

	\foreach \x in {0,2} {
		\foreach \y in {0,2} {
			\coordinate (\x\y) at (\x,\y);
			\path[draw, fill=black] (\x,\y) circle[radius=2pt] node {};
		}
	}

	\node at (00) [below left] {\scriptsize $0$};
	\node at (20) [below right] {\scriptsize $0$};
	\node at (02) [above left] {\scriptsize $\mu$};
	\node at (22) [above] {\scriptsize $\mu$};

	\draw[postaction={decorate}] (00) -- node[below] {\scriptsize $g_{ij}$} (20);
	\draw[postaction={decorate}] (02) -- node[auto] {\scriptsize $g_{ij}(\mu)$} (22);
	\draw[postaction={decorate}] (00) -- node[auto] {\rotatebox{90} {\scriptsize $\mathsf{hol}_{i\mu}$}} (02);
	\draw[postaction={decorate}] (20) -- node[right] {\rotatebox{90} {\scriptsize $\mathsf{hol}_{j\mu}$}} (22);

	\node at (1,1) {\rotatebox{-45}{\LARGE{}$\Rightarrow$}};
	\node at (1.3,1.3) {\scriptsize $g_{ij\mu}$};

\end{tikzpicture}
\end{center}

%% file: ColoredCube.tex
\begin{tikzpicture}[thin, decoration={
	markings,
	mark=at position 0.5 with {\arrow[black, line width=0.5mm]{stealth}}}
	]

	\foreach \x in {0,4} {
		\foreach \y in {0,4} {
			\foreach \z in {0,4} {
				\coordinate (\x\y\z) at (\x,\y,\z);
				\path[draw, fill=black] (\x,\y,\z) circle[radius=2pt] node {};
			}
		}
	}

	\node at (000) [left] {\scriptsize $\nu$};
	\node at (400) [right] {\scriptsize $\mu+\nu$};
	\node at (040) [above left] {\scriptsize $\nu$};
	\node at (004) [below left] {\scriptsize $0$};
	\node at (440) [above right] {\scriptsize $\mu+\nu$};
	\node at (044) [left] {\scriptsize $0$};
	\node at (404) [below right] {\scriptsize $\mu$};
	\node at (444) [right] {\scriptsize $\mu$};

	\draw[postaction={decorate}, dashed] (000) -- node[left] {\scriptsize \rotatebox{90} {$g_{ij}(\nu)$}} (040);
	\draw[postaction={decorate}] (040) -- node[above] {\scriptsize $\mathsf{hol}_{j\mu}(\nu)$} (440);
	\draw[postaction={decorate}] (400) -- node[right] {\scriptsize \rotatebox{90} {$g_{ij}(\mu+\nu)$}} (440);
	\draw[postaction={decorate}, dashed] (000) -- node[below] {\scriptsize $\mathsf{hol}_{i\mu}(\nu)$} (400);
	\draw[postaction={decorate}] (004) -- node[left] {\scriptsize \rotatebox{90} {$g_{ij}$}} (044);
	\draw[postaction={decorate}] (044) -- node[above] {\scriptsize $\mathsf{hol}_{j\mu}$} (444);
	\draw[postaction={decorate}] (404) -- node[right] {\scriptsize \rotatebox{90} {$g_{ij}(\mu)$}} (444);
	\draw[postaction={decorate}] (004) -- node[below] {\scriptsize $\mathsf{hol}_{i\mu}$} (404);
	\draw[postaction={decorate}] (044) -- node[left=-3pt] {\scriptsize \rotatebox{46} {$\mathsf{hol}_{j\nu}$}} (040);
	\draw[postaction={decorate}] (444) -- node[left=-6pt] {\scriptsize \rotatebox{46} {$\mathsf{hol}_{j\nu}(\mu)$}} (440);
	\draw[postaction={decorate}] (404) -- node[left=-8pt] {\scriptsize \rotatebox{46} {$\ \ \ \mathsf{hol}_{i\nu}(\mu)$}} (400);
	\draw[postaction={decorate}, dashed] (004) -- node[left=-10pt] {\scriptsize \rotatebox{46} {$\ \ \ \ \ \ \ \mathsf{hol}_{i\nu}$}} (000);

	\node at (0,1.8,2) {\rotatebox{120}{\LARGE{}$\Rightarrow$}};
	\node at (0,1.7,1.5) {\scriptsize \rotatebox{-60}{$g_{ij\nu}$}};

	\node at (2,2,4) {\rotatebox{135}{\LARGE{}$\Rightarrow$}};
	\node at (2.15,2.15,4) {\scriptsize \rotatebox{-45}{$g_{ij\mu}$}};

	\node at (4,1.8,2) {\rotatebox{120}{\LARGE{}$\Rightarrow$}};
	\node at (4,1.7,1.5) {\scriptsize \rotatebox{-60}{$g_{ij\nu}(\mu)$}};

	\node at (2,2,0) {\rotatebox{135}{\LARGE{}$\Rightarrow$}};
	\node at (2.15,2.15,0) {\scriptsize \rotatebox{-45}{$g_{ij\mu}(\nu)$}};

	\node at (2,0,2) {\rotatebox{145}{\LARGE{} $\Rightarrow$}};
	\node at (1.9,0,2.5) {\scriptsize \rotatebox{-35}{$\mathsf{hol}_{i\mu\nu}$}};

	\node at (2,4,2) {\rotatebox{145}{\LARGE{} $\Rightarrow$}};
	\node at (1.9,4,2.5) {\scriptsize \rotatebox{-35}{$\mathsf{hol}_{j\mu\nu}$}};
\end{tikzpicture}

%% file: cube_proof_9.tex
\begin{tikzpicture}[thin, decoration={
	markings,
	mark=at position 0.5 with {\arrow[black, line width=0.1mm]{stealth}}}
	]

	\foreach \x in {0,1} {
		\foreach \y in {0,1} {
			\foreach \z in {0,1} {
				\coordinate (\x\y\z) at (\x,\y,\z);
			}
		}
	}

	\draw[dotted] (000) -- (010);
	\draw[dotted] (010) -- (110);
	\draw[dotted] (100) -- (110);
	\draw[dotted] (000) -- (100);
	\draw[dotted] (001) -- (011);
	\draw[dotted] (011) -- (111);
	\draw[dotted] (101) -- (111);
	\draw[dotted] (001) -- (101);
	\draw[dotted] (011) -- (010);
	\draw[dotted] (111) -- (110);
	\draw[dotted] (101) -- (100);
	\draw[dotted] (001) -- (000);


	\path[draw, fill=black] (0,0,1) circle[radius=1pt] node {};		
	\path[draw, fill=black] (1,0,0) circle[radius=1pt] node {};		
	\path[draw, fill=black] (1,1,0) circle[radius=1pt] node {};		

	\draw[postaction={decorate}, shorten <=+4pt] (001) -- (101);
	\draw[postaction={decorate}, shorten >=+4pt] (101) -- (100);

	\draw[postaction={decorate}, shorten <=+4pt] (001) -- (000);
	\draw[postaction={decorate}, shorten >=+4pt] (000) -- (100);

	\draw[postaction={decorate}, shorten <=+4pt, shorten >=+4pt] (100) -- (110);

	\node at (0.5,0,0.5) {\rotatebox{145}{$\Rightarrow$}};


\end{tikzpicture}

%% file: cube_proof_10.tex
\begin{tikzpicture}[thin, decoration={
	markings,
	mark=at position 0.5 with {\arrow[black, line width=0.1mm]{stealth}}}
	]

	\foreach \x in {0,1} {
		\foreach \y in {0,1} {
			\foreach \z in {0,1} {
				\coordinate (\x\y\z) at (\x,\y,\z);
			}
		}
	}

	\draw[dotted] (000) -- (010);
	\draw[dotted] (010) -- (110);
	\draw[dotted] (100) -- (110);
	\draw[dotted] (000) -- (100);
	\draw[dotted] (001) -- (011);
	\draw[dotted] (011) -- (111);
	\draw[dotted] (101) -- (111);
	\draw[dotted] (001) -- (101);
	\draw[dotted] (011) -- (010);
	\draw[dotted] (111) -- (110);
	\draw[dotted] (101) -- (100);
	\draw[dotted] (001) -- (000);


	\path[draw, fill=black] (0,0,1) circle[radius=1pt] node {};		
	\path[draw, fill=black] (1,1,0) circle[radius=1pt] node {};		

	\draw[postaction={decorate}, shorten <=+4pt, shorten >=-1pt] (001) -- (101);
	\draw[postaction={decorate}, transform canvas={xshift=+1pt}, shorten >=+4pt] (100) -- (110);
	\draw[postaction={decorate}, transform canvas={xshift=+1pt}] (101) -- (100);

	\draw[postaction={decorate}, shorten <=+4pt] (001) -- (000);
	\draw[postaction={decorate}, shorten >=+4pt, shorten >=+1pt] (000) -- (100);
	\draw[postaction={decorate},, transform canvas={xshift=-1pt}, shorten >=+4pt] (100) -- (110);

	\node at (0.5,0,0.5) {\rotatebox{145}{$\Rightarrow$}};


\end{tikzpicture}

%% file: cube_proof_6.tex
\begin{tikzpicture}[thin, decoration={
	markings,
	mark=at position 0.5 with {\arrow[black, line width=0.1mm]{stealth}}}
	]

	\foreach \x in {0,1} {
		\foreach \y in {0,1} {
			\foreach \z in {0,1} {
				\coordinate (\x\y\z) at (\x,\y,\z);
			}
		}
	}

	\draw[dotted] (000) -- (010);
	\draw[dotted] (010) -- (110);
	\draw[dotted] (100) -- (110);
	\draw[dotted] (000) -- (100);
	\draw[dotted] (001) -- (011);
	\draw[dotted] (011) -- (111);
	\draw[dotted] (101) -- (111);
	\draw[dotted] (001) -- (101);
	\draw[dotted] (011) -- (010);
	\draw[dotted] (111) -- (110);
	\draw[dotted] (101) -- (100);
	\draw[dotted] (001) -- (000);


	\path[draw, fill=black] (0,0,1) circle[radius=1pt] node {};		
	\path[draw, fill=black] (1,0,1) circle[radius=1pt] node {};		
	\path[draw, fill=black] (1,1,0) circle[radius=1pt] node {};		

	\draw[postaction={decorate}, shorten <=+4pt] (101) -- (100);
	\draw[postaction={decorate}, shorten >=+4pt] (100) -- (110);

	\draw[postaction={decorate}, shorten <=+4pt] (101) -- (111);
	\draw[postaction={decorate}, shorten >=+4pt] (111) -- (110);

	\draw[postaction={decorate}, shorten <=+4pt, shorten >=+4pt] (001) -- (101);

	\node at (1,0.45,0.5) {\rotatebox{120}{$\Rightarrow$}};


\end{tikzpicture}

%% file: cube_proof_7.tex
\begin{tikzpicture}[thin, decoration={
	markings,
	mark=at position 0.5 with {\arrow[black, line width=0.1mm]{stealth}}}
	]

	\foreach \x in {0,1} {
		\foreach \y in {0,1} {
			\foreach \z in {0,1} {
				\coordinate (\x\y\z) at (\x,\y,\z);
			}
		}
	}

	\draw[dotted] (000) -- (010);
	\draw[dotted] (010) -- (110);
	\draw[dotted] (100) -- (110);
	\draw[dotted] (000) -- (100);
	\draw[dotted] (001) -- (011);
	\draw[dotted] (011) -- (111);
	\draw[dotted] (101) -- (111);
	\draw[dotted] (001) -- (101);
	\draw[dotted] (011) -- (010);
	\draw[dotted] (111) -- (110);
	\draw[dotted] (101) -- (100);
	\draw[dotted] (001) -- (000);


	\path[draw, fill=black] (0,0,1) circle[radius=1pt] node {};		
	\path[draw, fill=black] (1,1,0) circle[radius=1pt] node {};		

	\draw[postaction={decorate}, transform canvas={yshift=-1pt}, shorten <=+4pt] (001) -- (101);
	\draw[postaction={decorate}, transform canvas={yshift=-1pt}] (101) -- (100);
	\draw[postaction={decorate}, shorten <=-1pt, shorten >=+4pt] (100) -- (110);

	\draw[postaction={decorate}, transform canvas={yshift=+1pt}, shorten <=+4pt] (001) -- (101);
	\draw[postaction={decorate}, shorten <=+1pt] (101) -- (111);
	\draw[postaction={decorate}, shorten >=+4pt] (111) -- (110);

	\node at (1,0.45,0.5) {\rotatebox{120}{$\Rightarrow$}};


\end{tikzpicture}

%% file: cube_proof_3.tex
\begin{tikzpicture}[thin, decoration={
	markings,
	mark=at position 0.5 with {\arrow[black, line width=0.1mm]{stealth}}}
	]

	\foreach \x in {0,1} {
		\foreach \y in {0,1} {
			\foreach \z in {0,1} {
				\coordinate (\x\y\z) at (\x,\y,\z);
			}
		}
	}

	\draw[dotted] (000) -- (010);
	\draw[dotted] (010) -- (110);
	\draw[dotted] (100) -- (110);
	\draw[dotted] (000) -- (100);
	\draw[dotted] (001) -- (011);
	\draw[dotted] (011) -- (111);
	\draw[dotted] (101) -- (111);
	\draw[dotted] (001) -- (101);
	\draw[dotted] (011) -- (010);
	\draw[dotted] (111) -- (110);
	\draw[dotted] (101) -- (100);
	\draw[dotted] (001) -- (000);


	\path[draw, fill=black] (0,0,1) circle[radius=1pt] node {};		
	\path[draw, fill=black] (1,1,1) circle[radius=1pt] node {};		
	\path[draw, fill=black] (1,1,0) circle[radius=1pt] node {};		

	\draw[postaction={decorate}, shorten <=+4pt] (001) -- (101);
	\draw[postaction={decorate}, shorten >=+4pt] (101) -- (111);

	\draw[postaction={decorate}, shorten <=+4pt] (001) -- (011);
	\draw[postaction={decorate}, shorten >=+4pt] (011) -- (111);

	\draw[postaction={decorate}, shorten <=+4pt, shorten >=+4pt] (111) -- (110);

	\node at (0.5,0.5,1) {\rotatebox{135}{$\Rightarrow$}};


\end{tikzpicture}

%% file: cube_proof_4.tex
\begin{tikzpicture}[thin, decoration={
	markings,
	mark=at position 0.5 with {\arrow[black, line width=0.1mm]{stealth}}}
	]

	\foreach \x in {0,1} {
		\foreach \y in {0,1} {
			\foreach \z in {0,1} {
				\coordinate (\x\y\z) at (\x,\y,\z);
			}
		}
	}

	\draw[dotted] (000) -- (010);
	\draw[dotted] (010) -- (110);
	\draw[dotted] (100) -- (110);
	\draw[dotted] (000) -- (100);
	\draw[dotted] (001) -- (011);
	\draw[dotted] (011) -- (111);
	\draw[dotted] (101) -- (111);
	\draw[dotted] (001) -- (101);
	\draw[dotted] (011) -- (010);
	\draw[dotted] (111) -- (110);
	\draw[dotted] (101) -- (100);
	\draw[dotted] (001) -- (000);


	\path[draw, fill=black] (0,0,1) circle[radius=1pt] node {};		
	\path[draw, fill=black] (1,1,0) circle[radius=1pt] node {};		

	\draw[postaction={decorate}, shorten >=-1pt, shorten <=+4pt] (001) -- (101);
	\draw[postaction={decorate}, transform canvas={xshift=+1pt}] (101) -- (111);
	\draw[postaction={decorate}, transform canvas={xshift=-1pt}, shorten >=+4pt] (111) -- (110);

	\draw[postaction={decorate}, shorten <=+4pt] (001) -- (011);
	\draw[postaction={decorate}, shorten >=+4pt, shorten >=+1pt] (011) -- (111);
	\draw[postaction={decorate}, transform canvas={xshift=+1pt}, shorten >=+4pt] (111) -- (110);

	\node at (0.5,0.5,1) {\rotatebox{135}{$\Rightarrow$}};


\end{tikzpicture}

%% file: cube_proof_1.tex
\begin{tikzpicture}[thin, decoration={
	markings,
	mark=at position 0.5 with {\arrow[black, line width=0.1mm]{stealth}}}
	]

	\foreach \x in {0,1} {
		\foreach \y in {0,1} {
			\foreach \z in {0,1} {
				\coordinate (\x\y\z) at (\x,\y,\z);
			}
		}
	}

	\draw[dotted] (000) -- (010);
	\draw[dotted] (010) -- (110);
	\draw[dotted] (100) -- (110);
	\draw[dotted] (000) -- (100);
	\draw[dotted] (001) -- (011);
	\draw[dotted] (011) -- (111);
	\draw[dotted] (101) -- (111);
	\draw[dotted] (001) -- (101);
	\draw[dotted] (011) -- (010);
	\draw[dotted] (111) -- (110);
	\draw[dotted] (101) -- (100);
	\draw[dotted] (001) -- (000);


	\path[draw, fill=black] (0,0,1) circle[radius=1pt] node {};		
	\path[draw, fill=black] (0,1,1) circle[radius=1pt] node {};		
	\path[draw, fill=black] (1,1,0) circle[radius=1pt] node {};		

	\draw[postaction={decorate}, shorten <=+4pt] (011) -- (111);
	\draw[postaction={decorate}, shorten >=+4pt] (111) -- (110);

	\draw[postaction={decorate}, shorten <=+4pt] (011) -- (010);
	\draw[postaction={decorate}, shorten >=+4pt] (010) -- (110);

	\draw[postaction={decorate}, shorten <=+4pt, shorten >=+4pt] (001) -- (011);

	\node at (0.5,1,0.5) {\rotatebox{145}{$\Rightarrow$}};


\end{tikzpicture}

%% file: cube_proof_2.tex
\begin{tikzpicture}[thin, decoration={
	markings,
	mark=at position 0.5 with {\arrow[black, line width=0.1mm]{stealth}}}
	]

	\foreach \x in {0,1} {
		\foreach \y in {0,1} {
			\foreach \z in {0,1} {
				\coordinate (\x\y\z) at (\x,\y,\z);
			}
		}
	}

	\draw[dotted] (000) -- (010);
	\draw[dotted] (010) -- (110);
	\draw[dotted] (100) -- (110);
	\draw[dotted] (000) -- (100);
	\draw[dotted] (001) -- (011);
	\draw[dotted] (011) -- (111);
	\draw[dotted] (101) -- (111);
	\draw[dotted] (001) -- (101);
	\draw[dotted] (011) -- (010);
	\draw[dotted] (111) -- (110);
	\draw[dotted] (101) -- (100);
	\draw[dotted] (001) -- (000);


	\path[draw, fill=black] (0,0,1) circle[radius=1pt] node {};		
	\path[draw, fill=black] (1,1,0) circle[radius=1pt] node {};		

	\draw[postaction={decorate}, shorten <=+4pt, shorten <=+1pt] (011) -- (111);
	\draw[postaction={decorate}, shorten >=+4pt] (111) -- (110);
	\draw[postaction={decorate}, transform canvas={xshift=-1pt}, shorten <=+4pt] (001) -- (011);

	\draw[postaction={decorate}, shorten <=+4pt, shorten <=-0.5pt , transform canvas={xshift=-1pt}] (011) -- (010);
	\draw[postaction={decorate}, shorten >=+4pt, shorten <=-1pt] (010) -- (110);
	\draw[postaction={decorate}, transform canvas={xshift=+1pt}, shorten <=+4pt] (001) -- (011);

	\node at (0.5,1,0.5) {\rotatebox{145}{$\Rightarrow$}};


\end{tikzpicture}

%% file: cube_proof_13.tex
\begin{tikzpicture}[thin, decoration={
	markings,
	mark=at position 0.5 with {\arrow[black, line width=0.1mm]{stealth}}}
	]

	\foreach \x in {0,1} {
		\foreach \y in {0,1} {
			\foreach \z in {0,1} {
				\coordinate (\x\y\z) at (\x,\y,\z);
			}
		}
	}

	\draw[dotted] (000) -- (010);
	\draw[dotted] (010) -- (110);
	\draw[dotted] (100) -- (110);
	\draw[dotted] (000) -- (100);
	\draw[dotted] (001) -- (011);
	\draw[dotted] (011) -- (111);
	\draw[dotted] (101) -- (111);
	\draw[dotted] (001) -- (101);
	\draw[dotted] (011) -- (010);
	\draw[dotted] (111) -- (110);
	\draw[dotted] (101) -- (100);
	\draw[dotted] (001) -- (000);


	\path[draw, fill=black] (0,0,1) circle[radius=1pt] node {};		
	\path[draw, fill=black] (0,1,0) circle[radius=1pt] node {};		
	\path[draw, fill=black] (1,1,0) circle[radius=1pt] node {};		

	\draw[postaction={decorate}, shorten <=+4pt] (001) -- (000);
	\draw[postaction={decorate}, shorten >=+4pt] (000) -- (010);

	\draw[postaction={decorate}, shorten <=+4pt] (001) -- (011);
	\draw[postaction={decorate}, shorten >=+4pt] (011) -- (010);

	\draw[postaction={decorate}, shorten <=+4pt, shorten >=+4pt] (010) -- (110);

	\node at (0,0.45,0.5) {\rotatebox{120}{$\Rightarrow$}};


\end{tikzpicture}

%% file: cube_proof_14.tex
\begin{tikzpicture}[thin, decoration={
	markings,
	mark=at position 0.5 with {\arrow[black, line width=0.1mm]{stealth}}}
	]

	\foreach \x in {0,1} {
		\foreach \y in {0,1} {
			\foreach \z in {0,1} {
				\coordinate (\x\y\z) at (\x,\y,\z);
			}
		}
	}

	\draw[dotted] (000) -- (010);
	\draw[dotted] (010) -- (110);
	\draw[dotted] (100) -- (110);
	\draw[dotted] (000) -- (100);
	\draw[dotted] (001) -- (011);
	\draw[dotted] (011) -- (111);
	\draw[dotted] (101) -- (111);
	\draw[dotted] (001) -- (101);
	\draw[dotted] (011) -- (010);
	\draw[dotted] (111) -- (110);
	\draw[dotted] (101) -- (100);
	\draw[dotted] (001) -- (000);


	\path[draw, fill=black] (0,0,1) circle[radius=1pt] node {};		
	\path[draw, fill=black] (1,1,0) circle[radius=1pt] node {};		

	\draw[postaction={decorate}, shorten <=+4pt] (001) -- (000);
	\draw[postaction={decorate}, shorten >=+1pt] (000) -- (010);
	\draw[postaction={decorate}, shorten >=+4pt, transform canvas={yshift=-1pt}] (010) -- (110);

	\draw[postaction={decorate}, shorten <=+4pt, shorten >=-1pt] (001) -- (011);
	\draw[postaction={decorate}, transform canvas={yshift=+1pt}] (011) -- (010);
	\draw[postaction={decorate}, shorten >=+4pt, transform canvas={yshift=+1pt}] (010) -- (110);

	\node at (0,0.45,0.5) {\rotatebox{120}{$\Rightarrow$}};


\end{tikzpicture}

%% file: cube_proof_15.tex
\begin{tikzpicture}[thin, decoration={
	markings,
	mark=at position 0.5 with {\arrow[black, line width=0.1mm]{stealth}}}
	]

	\foreach \x in {0,1} {
		\foreach \y in {0,1} {
			\foreach \z in {0,1} {
				\coordinate (\x\y\z) at (\x,\y,\z);
			}
		}
	}

	\draw[dotted] (000) -- (010);
	\draw[dotted] (010) -- (110);
	\draw[dotted] (100) -- (110);
	\draw[dotted] (000) -- (100);
	\draw[dotted] (001) -- (011);
	\draw[dotted] (011) -- (111);
	\draw[dotted] (101) -- (111);
	\draw[dotted] (001) -- (101);
	\draw[dotted] (011) -- (010);
	\draw[dotted] (111) -- (110);
	\draw[dotted] (101) -- (100);
	\draw[dotted] (001) -- (000);


	\path[draw, fill=black] (0,0,1) circle[radius=1pt] node {};		
	\path[draw, fill=black] (1,1,0) circle[radius=1pt] node {};		

	\draw[postaction={decorate}, shorten <=+4pt] (001) -- (000);
	\draw[postaction={decorate}, shorten >=+1pt] (000) -- (010);
	\draw[postaction={decorate}, shorten >=+4pt, transform canvas={yshift=-1pt}] (010) -- (110);

	\draw[postaction={decorate}, shorten <=+4pt, shorten >=-1pt] (001) -- (011);
	\draw[postaction={decorate}, transform canvas={yshift=+1pt}] (011) -- (010);
	\draw[postaction={decorate}, shorten >=+4pt, transform canvas={yshift=+1pt}] (010) -- (110);

	\node at (0,0.45,0.5) {\rotatebox{-60}{$\Rightarrow$}};


\end{tikzpicture}

%% file: cube_proof_11.tex
\begin{tikzpicture}[thin, decoration={
	markings,
	mark=at position 0.5 with {\arrow[black, line width=0.1mm]{stealth}}}
	]

	\foreach \x in {0,1} {
		\foreach \y in {0,1} {
			\foreach \z in {0,1} {
				\coordinate (\x\y\z) at (\x,\y,\z);
			}
		}
	}

	\draw[dotted] (000) -- (010);
	\draw[dotted] (010) -- (110);
	\draw[dotted] (100) -- (110);
	\draw[dotted] (000) -- (100);
	\draw[dotted] (001) -- (011);
	\draw[dotted] (011) -- (111);
	\draw[dotted] (101) -- (111);
	\draw[dotted] (001) -- (101);
	\draw[dotted] (011) -- (010);
	\draw[dotted] (111) -- (110);
	\draw[dotted] (101) -- (100);
	\draw[dotted] (001) -- (000);


	\path[draw, fill=black] (0,0,1) circle[radius=1pt] node {};		
	\path[draw, fill=black] (0,0,0) circle[radius=1pt] node {};		
	\path[draw, fill=black] (1,1,0) circle[radius=1pt] node {};		

	\draw[postaction={decorate}, shorten <=+4pt] (000) -- (100);
	\draw[postaction={decorate}, shorten >=+4pt] (100) -- (110);

	\draw[postaction={decorate}, shorten <=+4pt] (000) -- (010);
	\draw[postaction={decorate}, shorten >=+4pt] (010) -- (110);

	\draw[postaction={decorate}, shorten <=+4pt, shorten >=+4pt] (001) -- (000);

	\node at (0.5,0.5,0) {\rotatebox{135}{$\Rightarrow$}};


\end{tikzpicture}

%% file: cube_proof_12.tex
\begin{tikzpicture}[thin, decoration={
	markings,
	mark=at position 0.5 with {\arrow[black, line width=0.1mm]{stealth}}}
	]

	\foreach \x in {0,1} {
		\foreach \y in {0,1} {
			\foreach \z in {0,1} {
				\coordinate (\x\y\z) at (\x,\y,\z);
			}
		}
	}

	\draw[dotted] (000) -- (010);
	\draw[dotted] (010) -- (110);
	\draw[dotted] (100) -- (110);
	\draw[dotted] (000) -- (100);
	\draw[dotted] (001) -- (011);
	\draw[dotted] (011) -- (111);
	\draw[dotted] (101) -- (111);
	\draw[dotted] (001) -- (101);
	\draw[dotted] (011) -- (010);
	\draw[dotted] (111) -- (110);
	\draw[dotted] (101) -- (100);
	\draw[dotted] (001) -- (000);


	\path[draw, fill=black] (0,0,1) circle[radius=1pt] node {};		
	\path[draw, fill=black] (1,1,0) circle[radius=1pt] node {};		

	\draw[postaction={decorate}, shorten <=+4pt, transform canvas={xshift=+1pt}] (001) -- (000);
	\draw[postaction={decorate}, shorten <=+1pt] (000) -- (100);
	\draw[postaction={decorate}, shorten >=+4pt] (100) -- (110);

	\draw[postaction={decorate}, shorten <=+4pt, transform canvas={xshift=-1pt}] (001) -- (000);
	\draw[postaction={decorate}, transform canvas={xshift=-1pt}] (000) -- (010);
	\draw[postaction={decorate}, shorten >=+4pt, shorten <=-1pt] (010) -- (110);

	\node at (0.5,0.5,0) {\rotatebox{135}{$\Rightarrow$}};


\end{tikzpicture}

%% file: cube_proof_12.5.tex
\begin{tikzpicture}[thin, decoration={
	markings,
	mark=at position 0.5 with {\arrow[black, line width=0.1mm]{stealth}}}
	]

	\foreach \x in {0,1} {
		\foreach \y in {0,1} {
			\foreach \z in {0,1} {
				\coordinate (\x\y\z) at (\x,\y,\z);
			}
		}
	}

	\draw[dotted] (000) -- (010);
	\draw[dotted] (010) -- (110);
	\draw[dotted] (100) -- (110);
	\draw[dotted] (000) -- (100);
	\draw[dotted] (001) -- (011);
	\draw[dotted] (011) -- (111);
	\draw[dotted] (101) -- (111);
	\draw[dotted] (001) -- (101);
	\draw[dotted] (011) -- (010);
	\draw[dotted] (111) -- (110);
	\draw[dotted] (101) -- (100);
	\draw[dotted] (001) -- (000);


	\path[draw, fill=black] (0,0,1) circle[radius=1pt] node {};		
	\path[draw, fill=black] (1,1,0) circle[radius=1pt] node {};		

	\draw[postaction={decorate}, shorten <=+4pt, transform canvas={xshift=+1pt}] (001) -- (000);
	\draw[postaction={decorate}, shorten <=+1pt] (000) -- (100);
	\draw[postaction={decorate}, shorten >=+4pt] (100) -- (110);

	\draw[postaction={decorate}, shorten <=+4pt, transform canvas={xshift=-1pt}] (001) -- (000);
	\draw[postaction={decorate}, transform canvas={xshift=-1pt}] (000) -- (010);
	\draw[postaction={decorate}, shorten >=+4pt, shorten <=-1pt] (010) -- (110);

	\node at (0.5,0.5,0) {\rotatebox{-45}{$\Rightarrow$}};


\end{tikzpicture}

%% file: ColoredPrism.tex
\begin{tikzpicture}[thin, decoration={
	markings,
	mark=at position 0.5 with {\arrow[black, line width=0.5mm]{stealth}}}
	]

	\foreach \x in {0,4} {
		\foreach \y in {0,4} {
			\coordinate (\x\y0) at (\x,\y,0);
			\path[draw, fill=black] (\x,\y,0) circle[radius=2pt] node {};
		}
	}
	\coordinate (404) at (4,0,4);
	\path[draw, fill=black] (4,0,4) circle[radius=2pt] node {};
	\coordinate (444) at (4,4,4);
	\path[draw, fill=black] (4,4,4) circle[radius=2pt] node {};

	\node at (040) [above left] {\scriptsize $0$};
	\node at (444) [below right] {\scriptsize $0$};
	\node at (440) [above right] {\scriptsize $0$};
	\node at (000) [below left] {\scriptsize $\mu$};
	\node at (404) [below right] {\scriptsize $\mu$};
	\node at (400) [below right] {\scriptsize $\mu$};

	\draw[postaction={decorate}] (040) -- node[above] {\scriptsize {$g_{ik}$}} (440);
	\draw[postaction={decorate}] (040) -- node[left] {\scriptsize \rotatebox{-30} {$g_{ij}$}} (444);
	\draw[postaction={decorate}] (444) -- node[right=-4pt] {\scriptsize \rotatebox{+45} {$g_{jk}$}} (440);

	\draw[postaction={decorate}, dashed] (000) -- node[above] {\scriptsize {$g_{ik}(\mu)$}} (400);
	\draw[postaction={decorate}] (000) -- node[left] {\scriptsize \rotatebox{-30} {$g_{ij}(\mu)$}} (404);
	\draw[postaction={decorate}] (404) -- node[right=-4pt] {\scriptsize \rotatebox{+45} {$g_{jk}(\mu)$}} (400);

	\draw[postaction={decorate}] (040) -- node[right] {\scriptsize \rotatebox{-90} {$\mathsf{hol}_{i\mu}$}} (000);
	\draw[postaction={decorate}] (444) -- node[right] {\scriptsize \rotatebox{-90} {$\mathsf{hol}_{j\mu}$}} (404);
	\draw[postaction={decorate}] (440) -- node[right] {\scriptsize \rotatebox{-90} {$\mathsf{hol}_{k\mu}$}} (400);

	\node at (2,2,0) {\rotatebox{45}{\LARGE{}$\Rightarrow$}};
	\node at (1.8,2.2,0) {\scriptsize \rotatebox{45}{$g_{ik\mu}$}};

	\node at (1.2,1,0) {\rotatebox{+45}{\LARGE{}$\Rightarrow$}};
	\node at (1,1.2,0) {\scriptsize \rotatebox{45}{$g_{ij\mu}$}};

	\node at (3.3,1.3,0) {\rotatebox{+75}{\LARGE{}$\Rightarrow$}};
	\node at (3.05,1.45,0) {\scriptsize \rotatebox{+75}{$g_{jk\mu}$}};

	\node at (2.3,3.4,0) {\rotatebox{+110}{\LARGE{}$\Rightarrow$}};
	\node at (2,3.3,0) {\scriptsize \rotatebox{+110}{$h_{ijk}$}};

	\node at (2.3,-0.7,0) {\rotatebox{+110}{\LARGE{}$\Rightarrow$}};
	\node at (1.9,-0.7,0) {\scriptsize \rotatebox{+110}{$h_{ijk}(\mu)$}};

\end{tikzpicture}

%% file: prism_proof_1.tex
\begin{tikzpicture}[thin, decoration={
	markings,
	mark=at position 0.5 with {\arrow[black, line width=0.1mm]{stealth}}}
	]

	\foreach \x in {0,1} {
		\foreach \y in {0,1} {
			\coordinate (\x\y0) at (\x,\y,0);
		}
	}

	\coordinate (404) at (1,0,1);
	\coordinate (444) at (1,1,1);

	\draw[dotted] (000) -- (100);
	\draw[dotted] (000) -- (101);
	\draw[dotted] (101) -- (100);
	\draw[dotted] (010) -- (110);
	\draw[dotted] (010) -- (111);
	\draw[dotted] (111) -- (110);
	\draw[dotted] (010) -- (000);
	\draw[dotted] (111) -- (101);
	\draw[dotted] (110) -- (100);


	\path[draw, fill=black] (0,1,0) circle[radius=1pt] node {};		
	\path[draw, fill=black] (1,0,0) circle[radius=1pt] node {};		

	\draw[postaction={decorate}, shorten <=+4pt] (010) -- (000);
	\draw[postaction={decorate}, shorten >=+4pt] (000) -- (100);

	\draw[postaction={decorate}, shorten <=+4pt] (010) -- (110);
	\draw[postaction={decorate}, shorten >=+4pt] (110) -- (100);

	\node at (0.5,0.5,0) {\rotatebox{45}{$\Rightarrow$}};

\end{tikzpicture}

%% file: prism_proof_2.tex
\begin{tikzpicture}[thin, decoration={
	markings,
	mark=at position 0.5 with {\arrow[black, line width=0.1mm]{stealth}}}
	]

	\foreach \x in {0,1} {
		\foreach \y in {0,1} {
			\coordinate (\x\y0) at (\x,\y,0);
		}
	}

	\coordinate (404) at (1,0,1);
	\coordinate (444) at (1,1,1);

	\draw[dotted] (000) -- (100);
	\draw[dotted] (000) -- (101);
	\draw[dotted] (101) -- (100);
	\draw[dotted] (010) -- (110);
	\draw[dotted] (010) -- (111);
	\draw[dotted] (111) -- (110);
	\draw[dotted] (010) -- (000);
	\draw[dotted] (111) -- (101);
	\draw[dotted] (110) -- (100);


	\path[draw, fill=black] (0,1,0) circle[radius=1pt] node {};		
	\path[draw, fill=black] (0,0,0) circle[radius=1pt] node {};		
	\path[draw, fill=black] (1,0,0) circle[radius=1pt] node {};		

	\draw[postaction={decorate}, shorten <=+4pt] (000) -- (101);
	\draw[postaction={decorate}, shorten >=+4pt] (101) -- (100);

	\draw[postaction={decorate}, shorten <=+4pt, shorten >=+4pt] (000) -- (100);

	\draw[postaction={decorate}, shorten <=+4pt, shorten >=+4pt] (010) -- (000);

	\node at (0.55,-0.14,0) {\scriptsize \rotatebox{+110}{$\Rightarrow$}};

\end{tikzpicture}

%% file: prism_proof_3.tex
\begin{tikzpicture}[thin, decoration={
	markings,
	mark=at position 0.5 with {\arrow[black, line width=0.1mm]{stealth}}}
	]

	\foreach \x in {0,1} {
		\foreach \y in {0,1} {
			\coordinate (\x\y0) at (\x,\y,0);
		}
	}

	\coordinate (404) at (1,0,1);
	\coordinate (444) at (1,1,1);

	\draw[dotted] (000) -- (100);
	\draw[dotted] (000) -- (101);
	\draw[dotted] (101) -- (100);
	\draw[dotted] (010) -- (110);
	\draw[dotted] (010) -- (111);
	\draw[dotted] (111) -- (110);
	\draw[dotted] (010) -- (000);
	\draw[dotted] (111) -- (101);
	\draw[dotted] (110) -- (100);


	\path[draw, fill=black] (0,1,0) circle[radius=1pt] node {};		
	\path[draw, fill=black] (1,0,0) circle[radius=1pt] node {};		

	\draw[postaction={decorate}, shorten <=-1pt] (000) -- (101);
	\draw[postaction={decorate}, shorten >=+4pt] (101) -- (100);

	\draw[postaction={decorate}, shorten <=+1pt, shorten >=+4pt, transform canvas={yshift=+1pt}] (000) -- (100);

	\draw[postaction={decorate}, shorten <=+4pt, shorten >=+1pt, transform canvas={xshift=+1pt}] (010) -- (000);
	\draw[postaction={decorate}, shorten <=+4pt, shorten >=+1pt, transform canvas={xshift=-1pt}] (010) -- (000);

	\node at (0.55,-0.14,0) {\scriptsize \rotatebox{+110}{$\Rightarrow$}};

\end{tikzpicture}

%% file: prism_proof_4.tex
\begin{tikzpicture}[thin, decoration={
	markings,
	mark=at position 0.5 with {\arrow[black, line width=0.1mm]{stealth}}}
	]

	\foreach \x in {0,1} {
		\foreach \y in {0,1} {
			\coordinate (\x\y0) at (\x,\y,0);
		}
	}

	\coordinate (404) at (1,0,1);
	\coordinate (444) at (1,1,1);

	\draw[dotted] (000) -- (100);
	\draw[dotted] (000) -- (101);
	\draw[dotted] (101) -- (100);
	\draw[dotted] (010) -- (110);
	\draw[dotted] (010) -- (111);
	\draw[dotted] (111) -- (110);
	\draw[dotted] (010) -- (000);
	\draw[dotted] (111) -- (101);
	\draw[dotted] (110) -- (100);


	\path[draw, fill=black] (0,1,0) circle[radius=1pt] node {};		
	\path[draw, fill=black] (1,0,0) circle[radius=1pt] node {};		

	\draw[postaction={decorate}, shorten <=-1pt] (000) -- (101);
	\draw[postaction={decorate}, shorten >=+4pt] (101) -- (100);

	\draw[postaction={decorate}, shorten <=+1pt, shorten >=+4pt, transform canvas={yshift=+1pt}] (000) -- (100);

	\draw[postaction={decorate}, shorten <=+4pt, shorten >=+1pt, transform canvas={xshift=+1pt}] (010) -- (000);
	\draw[postaction={decorate}, shorten <=+4pt, shorten >=+1pt, transform canvas={xshift=-1pt}] (010) -- (000);

	\node at (0.55,-0.14,0) {\scriptsize \rotatebox{-70}{$\Rightarrow$}};

\end{tikzpicture}

%% file: prism_proof_9.tex
\begin{tikzpicture}[thin, decoration={
	markings,
	mark=at position 0.5 with {\arrow[black, line width=0.1mm]{stealth}}}
	]

	\foreach \x in {0,1} {
		\foreach \y in {0,1} {
			\coordinate (\x\y0) at (\x,\y,0);
		}
	}

	\coordinate (404) at (1,0,1);
	\coordinate (444) at (1,1,1);

	\draw[dotted] (000) -- (100);
	\draw[dotted] (000) -- (101);
	\draw[dotted] (101) -- (100);
	\draw[dotted] (010) -- (110);
	\draw[dotted] (010) -- (111);
	\draw[dotted] (111) -- (110);
	\draw[dotted] (010) -- (000);
	\draw[dotted] (111) -- (101);
	\draw[dotted] (110) -- (100);


	\path[draw, fill=black] (0,1,0) circle[radius=1pt] node {};		
	\path[draw, fill=black] (1,0,1) circle[radius=1pt] node {};		
	\path[draw, fill=black] (1,0,0) circle[radius=1pt] node {};		

	\draw[postaction={decorate}, shorten <=+4pt] (010) -- (111);
	\draw[postaction={decorate}, shorten >=+4pt] (111) -- (101);

	\draw[postaction={decorate}, shorten <=+4pt] (010) -- (000);
	\draw[postaction={decorate}, shorten >=+4pt] (000) -- (101);

	\draw[postaction={decorate}, shorten <=+4pt, shorten >=+4pt] (101) -- (100);

	\node at (0.3,0.25,0) {\rotatebox{+45}{$\Rightarrow$}};

\end{tikzpicture}

%% file: prism_proof_10.tex
\begin{tikzpicture}[thin, decoration={
	markings,
	mark=at position 0.5 with {\arrow[black, line width=0.1mm]{stealth}}}
	]

	\foreach \x in {0,1} {
		\foreach \y in {0,1} {
			\coordinate (\x\y0) at (\x,\y,0);
		}
	}

	\coordinate (404) at (1,0,1);
	\coordinate (444) at (1,1,1);

	\draw[dotted] (000) -- (100);
	\draw[dotted] (000) -- (101);
	\draw[dotted] (101) -- (100);
	\draw[dotted] (010) -- (110);
	\draw[dotted] (010) -- (111);
	\draw[dotted] (111) -- (110);
	\draw[dotted] (010) -- (000);
	\draw[dotted] (111) -- (101);
	\draw[dotted] (110) -- (100);


	\path[draw, fill=black] (0,1,0) circle[radius=1pt] node {};		
	\path[draw, fill=black] (1,0,0) circle[radius=1pt] node {};		

	\draw[postaction={decorate}, shorten <=+4pt] (010) -- (111);
	\draw[postaction={decorate}, shorten >=+1pt] (111) -- (101);

	\draw[postaction={decorate}, shorten <=+4pt] (010) -- (000);
	\draw[postaction={decorate}, shorten >=-1pt] (000) -- (101);

	\draw[postaction={decorate}, shorten >=+4pt, transform canvas={yshift=+1pt}] (101) -- (100);
	\draw[postaction={decorate}, shorten >=+4pt, shorten <=+0.5pt, transform canvas={yshift=-1pt}] (101) -- (100);

	\node at (0.3,0.25,0) {\rotatebox{+45}{$\Rightarrow$}};

\end{tikzpicture}

%% file: prism_proof_5.tex
\begin{tikzpicture}[thin, decoration={
	markings,
	mark=at position 0.5 with {\arrow[black, line width=0.1mm]{stealth}}}
	]

	\foreach \x in {0,1} {
		\foreach \y in {0,1} {
			\coordinate (\x\y0) at (\x,\y,0);
		}
	}

	\coordinate (404) at (1,0,1);
	\coordinate (444) at (1,1,1);

	\draw[dotted] (000) -- (100);
	\draw[dotted] (000) -- (101);
	\draw[dotted] (101) -- (100);
	\draw[dotted] (010) -- (110);
	\draw[dotted] (010) -- (111);
	\draw[dotted] (111) -- (110);
	\draw[dotted] (010) -- (000);
	\draw[dotted] (111) -- (101);
	\draw[dotted] (110) -- (100);


	\path[draw, fill=black] (0,1,0) circle[radius=1pt] node {};		
	\path[draw, fill=black] (1,1,1) circle[radius=1pt] node {};		
	\path[draw, fill=black] (1,0,0) circle[radius=1pt] node {};		

	\draw[postaction={decorate}, shorten <=+4pt] (111) -- (101);
	\draw[postaction={decorate}, shorten >=+4pt] (101) -- (100);

	\draw[postaction={decorate}, shorten <=+4pt] (111) -- (110);
	\draw[postaction={decorate}, shorten >=+4pt] (110) -- (100);

	\draw[postaction={decorate}, shorten <=+4pt, shorten >=+4pt] (010) -- (111);

	\node at (0.825,0.325,0) {\rotatebox{+75}{$\Rightarrow$}};

\end{tikzpicture}

%% file: prism_proof_6.tex
\begin{tikzpicture}[thin, decoration={
	markings,
	mark=at position 0.5 with {\arrow[black, line width=0.1mm]{stealth}}}
	]

	\foreach \x in {0,1} {
		\foreach \y in {0,1} {
			\coordinate (\x\y0) at (\x,\y,0);
		}
	}

	\coordinate (404) at (1,0,1);
	\coordinate (444) at (1,1,1);

	\draw[dotted] (000) -- (100);
	\draw[dotted] (000) -- (101);
	\draw[dotted] (101) -- (100);
	\draw[dotted] (010) -- (110);
	\draw[dotted] (010) -- (111);
	\draw[dotted] (111) -- (110);
	\draw[dotted] (010) -- (000);
	\draw[dotted] (111) -- (101);
	\draw[dotted] (110) -- (100);


	\path[draw, fill=black] (0,1,0) circle[radius=1pt] node {};		
	\path[draw, fill=black] (1,0,0) circle[radius=1pt] node {};		

	\draw[postaction={decorate}, shorten <=+1pt] (111) -- (101);
	\draw[postaction={decorate}, shorten >=+4pt] (101) -- (100);

	\draw[postaction={decorate}, shorten <=+1pt] (111) -- (110);
	\draw[postaction={decorate}, shorten >=+4pt] (110) -- (100);

	\draw[postaction={decorate}, shorten <=+4pt, shorten >=-0.8pt, transform canvas={yshift=+1pt}] (010) -- (111);
	\draw[postaction={decorate}, shorten <=+4pt, transform canvas={yshift=-1pt}] (010) -- (111);

	\node at (0.825,0.325,0) {\rotatebox{+75}{$\Rightarrow$}};

\end{tikzpicture}

%% file: prism_proof_7.tex
\begin{tikzpicture}[thin, decoration={
	markings,
	mark=at position 0.5 with {\arrow[black, line width=0.1mm]{stealth}}}
	]

	\foreach \x in {0,1} {
		\foreach \y in {0,1} {
			\coordinate (\x\y0) at (\x,\y,0);
		}
	}

	\coordinate (404) at (1,0,1);
	\coordinate (444) at (1,1,1);

	\draw[dotted] (000) -- (100);
	\draw[dotted] (000) -- (101);
	\draw[dotted] (101) -- (100);
	\draw[dotted] (010) -- (110);
	\draw[dotted] (010) -- (111);
	\draw[dotted] (111) -- (110);
	\draw[dotted] (010) -- (000);
	\draw[dotted] (111) -- (101);
	\draw[dotted] (110) -- (100);


	\path[draw, fill=black] (0,1,0) circle[radius=1pt] node {};		
	\path[draw, fill=black] (1,1,0) circle[radius=1pt] node {};		
	\path[draw, fill=black] (1,0,0) circle[radius=1pt] node {};		

	\draw[postaction={decorate}, shorten <=+4pt] (010) -- (111);
	\draw[postaction={decorate}, shorten >=+4pt] (111) -- (110);

	\draw[postaction={decorate}, shorten <=+4pt, shorten >=+4pt] (010) -- (110);

	\draw[postaction={decorate}, shorten <=+4pt, shorten >=+4pt] (110) -- (100);

	\node at (0.575,0.85,0) {\scriptsize \rotatebox{+110}{$\Rightarrow$}};

\end{tikzpicture}

%% file: prism_proof_8.tex
\begin{tikzpicture}[thin, decoration={
	markings,
	mark=at position 0.5 with {\arrow[black, line width=0.1mm]{stealth}}}
	]

	\foreach \x in {0,1} {
		\foreach \y in {0,1} {
			\coordinate (\x\y0) at (\x,\y,0);
		}
	}

	\coordinate (404) at (1,0,1);
	\coordinate (444) at (1,1,1);

	\draw[dotted] (000) -- (100);
	\draw[dotted] (000) -- (101);
	\draw[dotted] (101) -- (100);
	\draw[dotted] (010) -- (110);
	\draw[dotted] (010) -- (111);
	\draw[dotted] (111) -- (110);
	\draw[dotted] (010) -- (000);
	\draw[dotted] (111) -- (101);
	\draw[dotted] (110) -- (100);


	\path[draw, fill=black] (0,1,0) circle[radius=1pt] node {};		
	\path[draw, fill=black] (1,0,0) circle[radius=1pt] node {};		

	\draw[postaction={decorate}, shorten <=+4pt, shorten >=-1pt] (010) -- (111);
	\draw[postaction={decorate}, shorten >=+3pt, shorten <=-1pt, transform canvas={xshift=+1pt}] (111) -- (110);

	\draw[postaction={decorate}, shorten <=+4pt, shorten >=-1pt] (010) -- (110);

	\draw[postaction={decorate}, shorten >=+4pt, shorten <=+2pt, transform canvas={xshift=-1pt}] (110) -- (100);
	\draw[postaction={decorate}, shorten >=+4pt, transform canvas={xshift=+1pt}] (110) -- (100);

	\node at (0.575,0.85,0) {\scriptsize \rotatebox{+110}{$\Rightarrow$}};

\end{tikzpicture}